# Dielectric Hysteresis, Relaxation Dynamics, and Non-volatile Memory Effect in Carbon Nanotube Dispersed Liquid Crystal

Rajratan Basu and Germano S. Iannacchione

*Order-Disorder Phenomena Laboratory, Department of Physics, Worcester Polytechnic Institute, Worcester, MA 01609, USA*

## Abstract

Self-organizing nematic liquid crystals (LC) impart their orientational order onto dispersed carbon nanotubes (CNTs) and obtain CNT-self-assembly on a macroscopic dimension. The nanotubes-long axis, being coupled to the nematic director, enables orientational manipulation via the LC nematic reorientation. Electric field induced director rotation of a nematic LC+CNT system is of potential interest due to its possible application as a nano-electromechanical system. Electric field and temperature dependence of dielectric properties of an LC+CNT composite system have been investigated to understand the principles governing CNT-assembly mediated by the LC. In the LC+CNT nematic phase, the dielectric relaxation on removing the applied field follows a single exponential decay, exhibiting a faster decay response than the pure LC above a threshold field. The observed dielectric hysteresis on field-cycling in the nematic phase for the composite indicates an improvement in the spontaneous polarization in the nematic matrix due to the dispersed CNTs. Observations in the isotropic phase coherently combine to confirm the presence of anisotropic *pseudo-nematic* domains stabilized by the LC – CNT anchoring energy. These polarized domains maintain local directors, and respond to external fields, but, do not relax back to the original state on switching the field off, showing non-volatile memory effect.

## I. INTRODUCTION

Self-organizing nematic liquid crystal (LC) has gained interest in recent years for transferring orientational order onto suspended nanoparticles [1,2,3,4,5,6,7,8]. It has been demonstrated that inside a nematic liquid crystal matrix, the long axes of carbon nanotubes (CNTs) orient parallel to the *director field* (average direction of LC molecules) with an orientational order parameter $S$ between 0.6 to 0.9 [1,2,7], while bulk nematic liquid crystals themselves have orientational order of $S \approx 0.6$. Recent theoretical work shows that a strong interaction, mainly due to surface anchoring with a binding energy of about –2 eV for $\pi$–$\pi$ stacking between LC–CNT [4,9], is associated with the CNT alignment mechanism in the nematic state. It has also been known theoretically for some time that the distribution of orientations of the nano-size anisotropic guest particles in an anisotropic and ordered solution is along the symmetry axis of the solution [10]. An anisotropic nematic LC shows cylindrical symmetry along the director field [10,11]. Thus, the anchoring energy favors the dispersed CNT long-axis parallel to the director field, minimizing the elastic distortion of the nematic matrix, which is essential a minimization of excluded volume [10]. This is shown schematically in Fig. 1.

As the CNTs are much thinner than the elastic penetration length, the alignment is driven by the coupling of the unperturbed director field to the anisotropic interfacial tension of the CNTs in the nematic LC matrix [12]. Thus, the concentration of CNTs in LC is a very important parameter for this alignment process as mono-dispersion without any agglomerates is need. The dilute suspensions (as low as 0.005 wt%) are stable because well dispersed CNTs individually (not in bundles) do not perturb the director field significantly. Consequently, the nanotubes share their intrinsic properties with the LC matrix, such as electrical conductivity [2], due to the alignment with the LC molecules. Thus, comprehensive understanding of the interaction of CNTs with an LC and the principles governing their self-assembly through an LC mediated interactions is an important and active area of research. Exploiting the nematic LC for nanotemplating purposes and controlling the director by applying fields make the LC+CNT mixture an attractive anisotropic physical system to study the Fréedericksz switching through an electromechanical response at the nanoscale level. This electromechanical switching behavior of CNTs embedded in a nematic platform enables the possibilities of using CNTs in micro/nano electronics for molecular wiring. After a field-induced director rotation of the nematic in an LC+CNT [1,2,3] contained in a planar LC cell [13], the LC molecules as well as the CNTs dynamically reorient back into their original orientation on the immediate removal of the electric field, exhibiting the intrinsic relaxation dynamics. The LC media, such as nematic phase or isotropic phase, strongly influences the reorientation and relaxation mechanisms of CNTs.

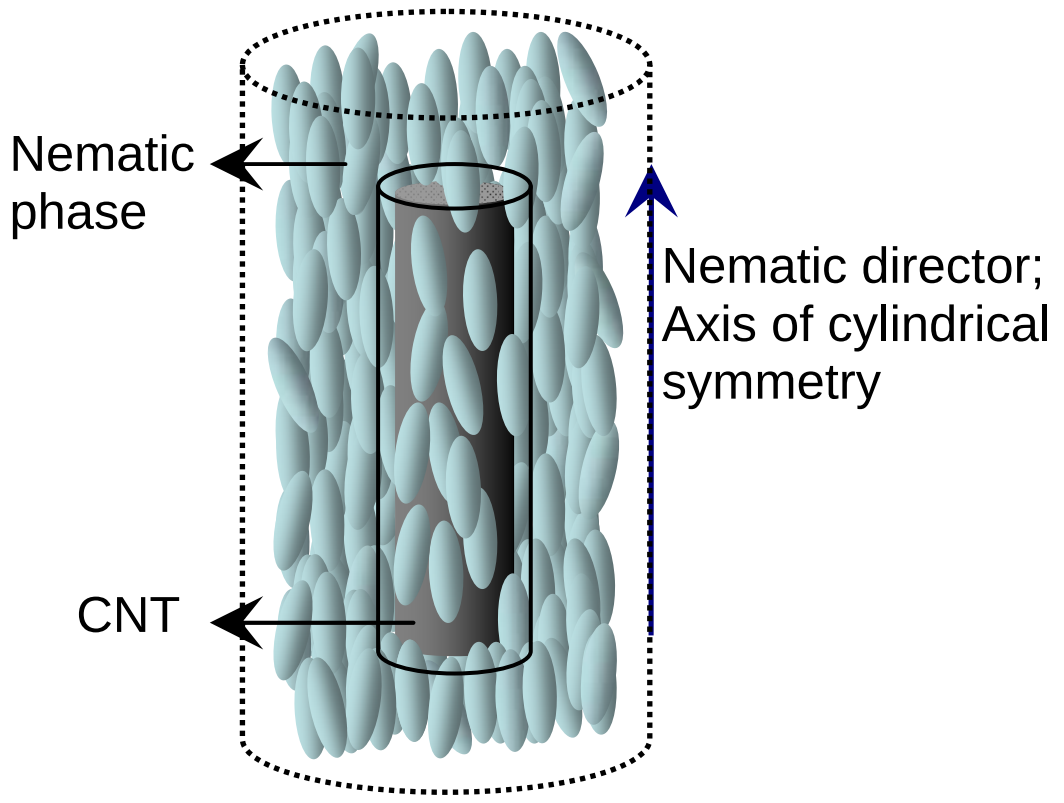

**FIGURE 1:** Schematic diagram of symmetry matching; Dotted cylinder shows the cylindrical symmetry of uniaxial nematic phase; the small cylinder shows the cylindrical confinement of a CNT.

In this paper, we report the dynamic response, the hysteresis effect, and the temperature dependence of the average dielectric constant ( $\bar{\varepsilon}$ ) for multiwall carbon nanotubes (MWCNTs) dispersed in 4-cyano-4'-pentylbiphenyl (5CB) LC in both the nematic and isotropic phases. The nematic phase shows dielectric anisotropy due to the anisotropic nature of the LC molecules where $\varepsilon_{\parallel}$ and $\varepsilon_{\perp}$ are the components parallel and perpendicular to the molecular long axis, respectively. For a positive dielectric anisotropic LC, $\varepsilon_{\parallel} > \varepsilon_{\perp}$, and so, the director field reorients parallel to an applied electric field. In a uniform homogeneously aligned parallel-plate cell configuration, the nematic director is aligned perpendicular to the applied electric field due to surface anchoring but the director can reorient parallel to the field if the field magnitude is above some critical threshold. This is the essence of a Fréedericksz transition and an ac-capacitive measurement of the $\bar{\varepsilon}$ will reveal $\varepsilon_{\perp}$ below and $\varepsilon_{\parallel}$ above this switching, the exact values depending on frequency. Having very high aspect ration, CNTs also exhibit dielectric anisotropy.

Following this introduction, a description of the materials, sample preparation and, ac capacitive bridge technique are given in Sec. II. Dielectric hysteresis, dynamic response of $\bar{\varepsilon}$, and temperature dependence of $\bar{\varepsilon}$ are presented in Sec. III, followed by conclusions in Sec. IV.

## II. EXPERIMENTAL PROCEDURES

### A. Materials and sample preparation

The MWCNT sample used for this experiments contains nanotubes < 8 nm in diameter and 0.5 – 2 μm in length with purity 95%, purchased from *Nanostructured & Amorphous Materials, Inc*. The well characterized LC 5CB used for this experiments has nematic to isotropic phase transition at temperature $T_{IN}$ = 35°C. A small amount (0.005 wt %) of MWCNT sample was dispersed in 5CB and the mixture was ultrasonicated for 5 hours to reduce the bundling tendency of CNTs. Soon after ultrasonication, the mixture was degassed under vacuum at 40°C for at least two hours. The mixture then was filled into a homogeneous LC cell (5 × 5 mm$^2$ indium tin oxide (ITO) coated area and 20 μm spacing) by capillary action, housed in a temperature controlled bath. The cell spacing filters out any nanotube aggregates larger than the spacing dimension. Surface treatment inside the LC cell induces the planar alignment to the nematic director [13]. Empty LC cells were measured separately first in order to extract the $\bar{\varepsilon}$. The relaxation dynamics also depend on cell configuration; for comparisons, the same type of cells was used for both pure 5CB and 5CB+MWCNT.

### B. AC-capacitance bridge technique and dielectric spectrometer

The dielectric measurements were performed by the ac capacitance bridge technique [14,15,16], operating with a probing field $E$ at 100 kHz frequency. Comparison between the empty and sample filled cell capacitance allows for an relative measurement of dielectric constant $\bar{\varepsilon}$ (100 kHz) with respect to the empty cell. The probing field, $E$ (100 kHz), in a capacitive measurement, is in phase and at the same frequency as the measurement of the complex dielectric constant, $\bar{\varepsilon}$ (100 kHz). Thus, in the complex *rotating-frame* of the measurement, the probing field $E$ (100 kHz) can be considered a 'static' field [17]. After the sample was freshly loaded into the cell, dielectric hysteresis experiments were performed by cycling the $E$ (100 kHz) field magnitude ranging from 10 ↔ 250 kV/m and monitoring $\bar{\varepsilon}$ ($E$), as shown in Fig. 2. The field annealing treatment trains the sample to improve the nematic ordering controlling any defects in the nematic matrix. Temperature annealing was also performed by heating the system to isotropic phase and then cooling it down to the nematic phase; but, no thermal hysteresis was observed.

After the completion of field and temperature annealing, an external ac electric field pulse, $E_{ac}$ (1 MHz) of 30 seconds duration was applied across the cell at magnitudes ranging from 0 to 250 kV/m. The field $E_{ac}$ (1 MHz) is independent of the capacitance bridge and measurement technique. The reason for applying the ac field (not dc) is to avoid the affect of ion migration or ionic conduction on the dielectric relaxation measurements. Once $E_{ac}$ was turned off (at $t$ = 0 sec), isothermal average dielectric $\bar{\varepsilon}$ ($t$) measurements were carried in the nematic ($T$ = 25° C) and isotropic ($T$ = 37° C) phases as a function of time. See Fig. 3 and Fig. 6. The magnitude of the probing field $E$ (5 kV/m, 100 kHz) was kept far below the Freedericksz reorientation threshold field during this measurement. The LC 5CB does not exhibit any tumbling relaxation mode [18] and MWCNTs show no space charge or dipole orientation dynamics at this probing frequency, 100 kHz [16,19]. Therefore, the observed dielectric relaxation is caused mainly by a mechanical relaxation mechanism of the director on turning $E_{ac}$ off. Temperature dependence of $\bar{\varepsilon}$ was studied, shown in Fig. 8, in order to understand the behavior of nematic and isotropic phase of the LC in presence of CNTs. Temperature dependence of $\bar{\varepsilon}$ provides information about isotropic to nematic phase transition and structural changes due to the anchoring energy arising from the presence of CNTs.

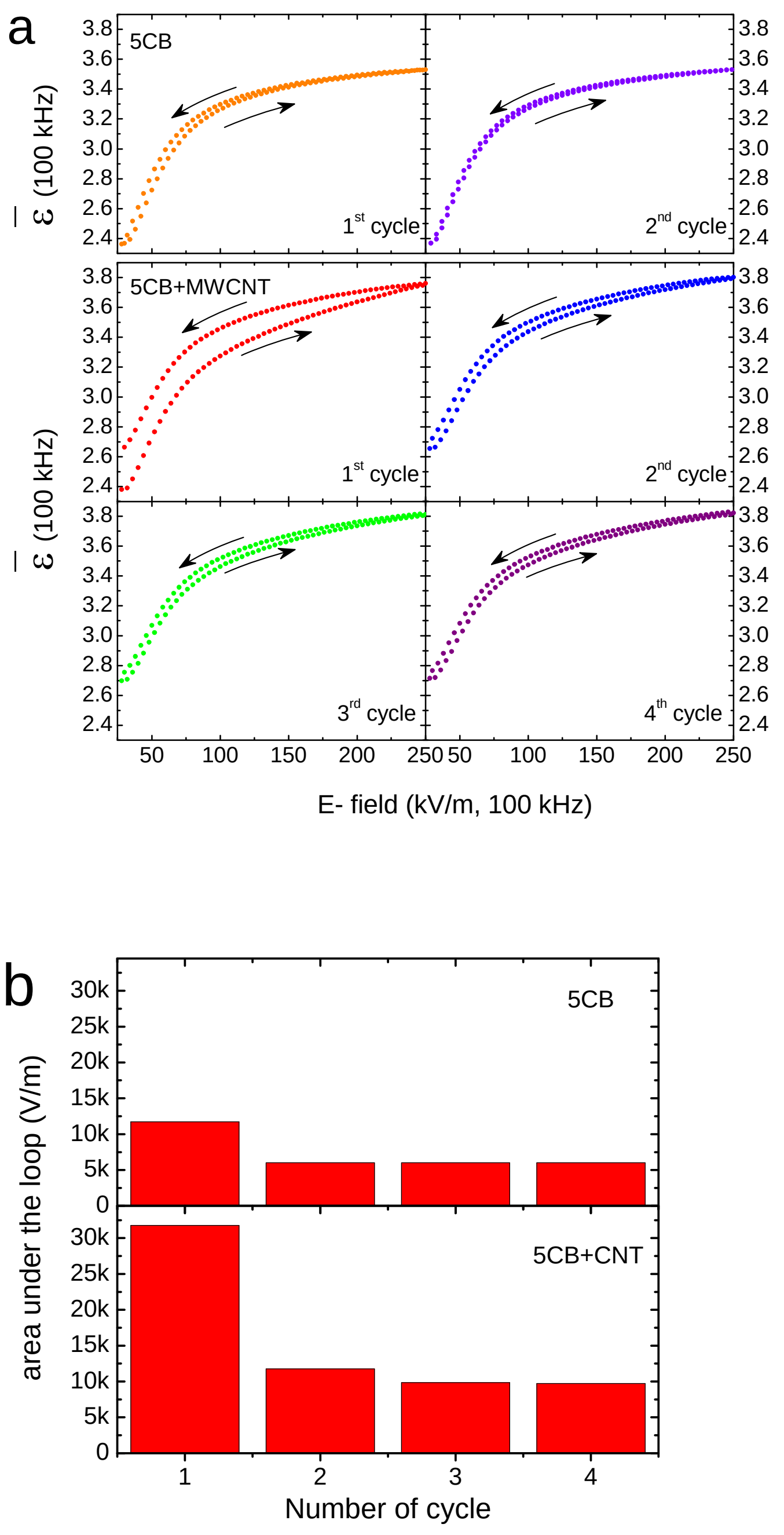

**FIGURE 2:** **a)** Field annealing hysteresis cycles, $\bar{\varepsilon}$ vs. $\boldsymbol{E}$, for 5CB (top 2 panels) and 5CB+MWCNT (bottom 4 panels) in the nematic phase ($T$ = 25°C). **b)** The area under the hysteresis loop as a function number of cycle for 5CB and 5CB+MWCNT.

## III. RESULTS AND DISCUSSIONS

### A. Dielectric hysteresis in the nematic phase

Figure 2a depicts the dielectric hysteresis effects for 5CB and 5CB+MWCNT in the nematic phase ($T$ = 25°C). As an ac field is used for this annealing process, ion migration or ionic conduction has no contribution to the dielectric behavior observed. Thus, the hysteresis effects observed is not caused by the residual dc effect. The area under the hysteresis loop, which is proportional to the energy lost during the cycle, is shown in Fig. 2b. The hysteresis area for bulk 5CB decreases by a considerable amount after the first cycle and then remains constant for rest of the number of cycles performed. A vivid change in the dielectric hysteresis has been observed for the composite system. The first cycle shows the highest amount the of hysteresis loss which gradually decreases through the 2nd and 3rd cycle, reaching saturation at the 4th cycle. After the 4th cycle the hysteresis area remains the same through out the rest of the number of annealing cycles performed. It is important to note that this dramatic change is observed due to the presence of only 0.005 wt% of MWCNT in the nematic LC media. This ferroelectric-type hysteresis effect clearly indicates that the spontaneous polarization in the nematic LC system increases by a considerable amount due to the presence of a small amount of CNTs, suggesting a structural modification in the nematic matrix. Individual and isolated CNTs can induce alignment on the director field along their long axis due to LC–CNT surface anchoring [6]. So, it is possible that LC-CNT coupling improves the nematic order. This causes an increment in the spontaneous polarization in the system, showing a ferroelectric-type hysteresis effect. The hysteresis area has been found to be increasing with increasing CNT concentration. Also, the multiple field-annealing controls LC defects [20] and perhaps results in a reduction of the defects in the matrix inside the cell - hence, decreasing hysteresis area with increasing cycle number for the first few cycles.

### B. Relaxation dynamics of $\bar{\varepsilon}$ in the nematic phase

The planar rubbing direction on the surface of the electrodes inside the LC cell acts as an anchoring field that induces homogeneous alignment on the first few LC layers touching the top and bottom electrodes. Then, the elastic interaction between the LC molecules makes the homogeneous alignment propagate through whole media, obtaining a planar director profile inside the cell. Being embedded in the nematic matrix CNT long axis also follows the planar director field. Electric field induced director reorientation occurs when the torques, due to the external electric field, overcome the elastic interactions between LC molecules and, through surface coupling, the CNT long axis follows the director rotation. Soon after the field goes off, these restoring forces, between the planar surface state and LC director, drive the system back to the planar configuration through a mechanical rotation.

The dielectric constant $\bar{\varepsilon}$ as a function of time and $E_{ac}$, after switching $E_{ac}$ off for the 5CB+MWCNT sample in the nematic phase ($T$ = 25° C), is shown in Fig. 3. The relaxation of $\bar{\varepsilon}$ follows a single-exponential decay, reaching to its original value. The field-saturated dielectric used as a nano-electromechanical system and micro/nano switch exploiting the high electrical conductivity of CNTs along the long axis. It is important to point out that the $E_{ac}$ driven director rotation is a much faster response than the relaxation response (decay) on switching $E_{ac}$ off. On switching $E_{ac}$ on and off this system acts as a *nano-oscillator* governing two different characteristic frequencies.

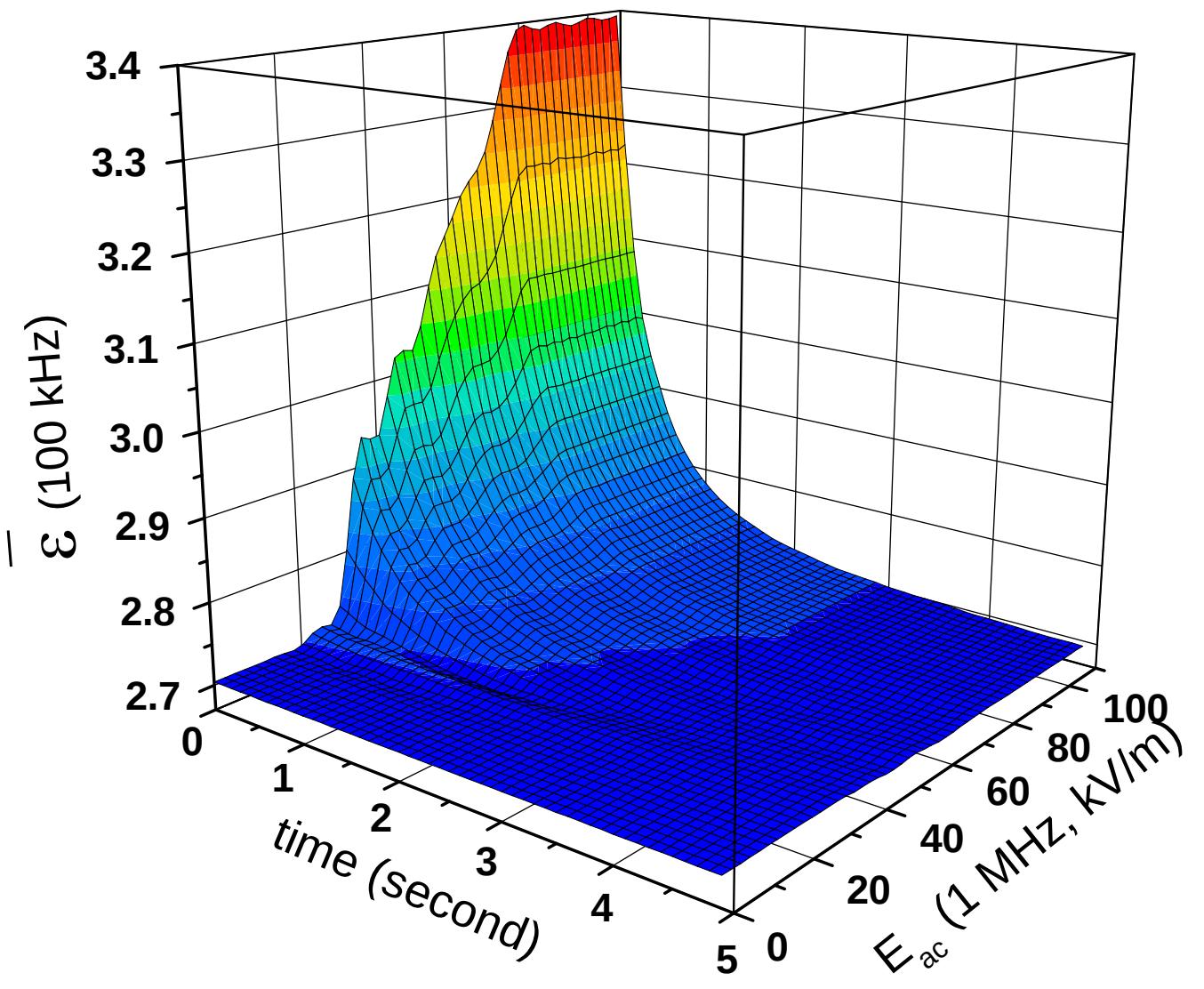

**FIGURE 3:** Dynamic response of the average dielectric constant $\bar{\varepsilon}$ for 5CB+MWCNT as a function of *time* and $E_{ac}$ (1 MHz) in the nematic phase ($T$ = 25°C) after $E_{ac}$ goes off.

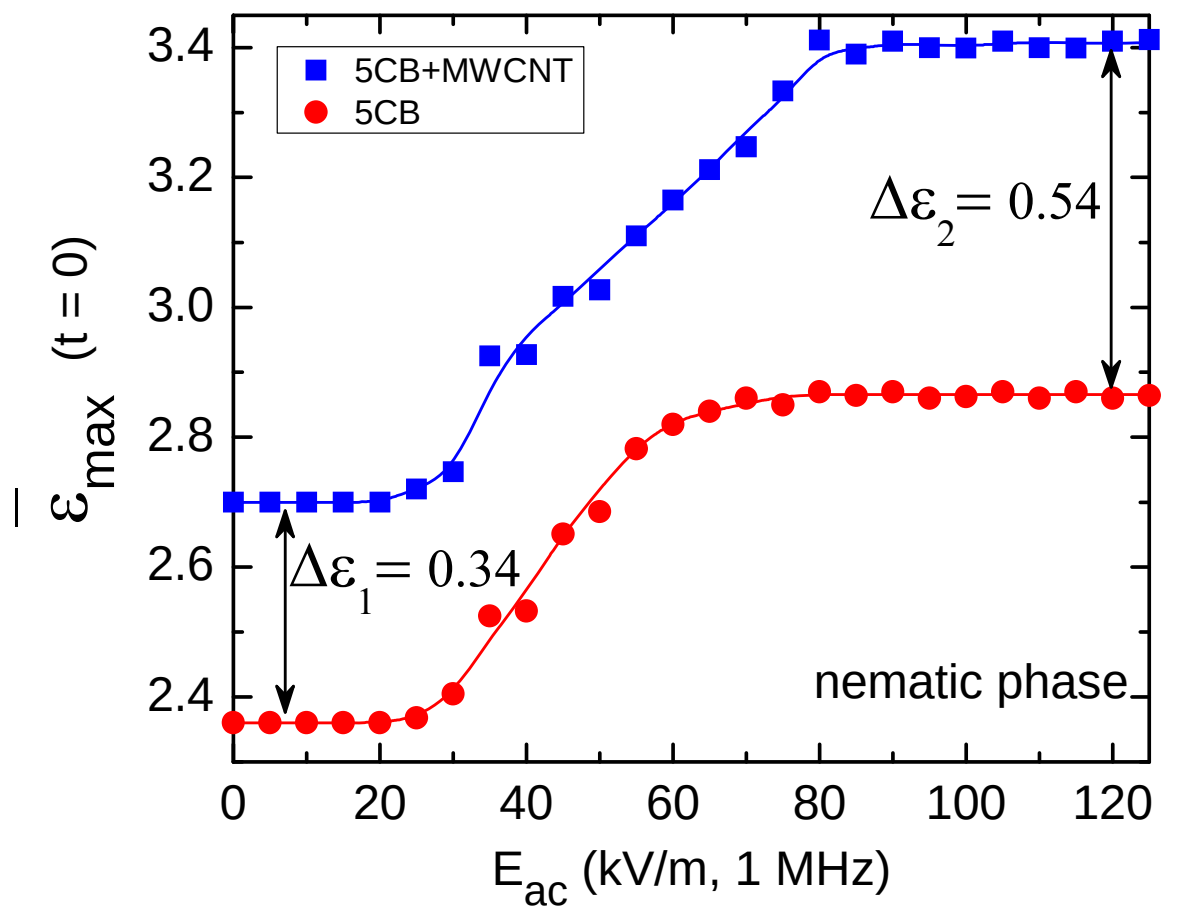

**FIGURE 4:** Field-saturated dielectric constant, $\bar{\varepsilon}_{max}$ ( $\bar{\varepsilon}$ at $t$ = 0) as a function of $E_{ac}$ for 5CB and 5CB+MWCNT in the nematic phase ($T$ = 25°C). Lines represent guide to the eye.

constant, $\bar{\varepsilon}_{max}$ ( $\bar{\varepsilon}$ at $t = 0$, from Fig 3) for each relaxation is plotted as a function of $E_{ac}$ in Fig. 4 and is directly associated with the director profile. The value of $\bar{\varepsilon}_{max}$ starts to increase above $E_{ac}$ = 20 kV/m for both pure LC and LC+CNT samples, confirming the director reorientation from planar to homeotropic, but, $\bar{\varepsilon}_{max}$ saturates at a higher field for the composite sample than pure 5CB. See Fig. 4. This is probably due to the higher aspect ratio of CNTs that require higher fields to fully reorient. As mentioned earlier, because of having very high aspect ration, CNTs also exhibit dielectric anisotropy and contribute their $\varepsilon_{\perp}$ or $\varepsilon_{\parallel}$ to the system depending on a particular orientation. For the homogeneous director profile ($E_{ac}$ < 20 kV/m in Fig. 4), individual CNTs being perpendicular to the measuring field, contribute their average $\varepsilon_{\perp}$ to the average dielectric constant of the system. Figure 4 shows that the average dielectric constant of the composite system increases by an amount $\Delta\varepsilon_1 = 0.34$. After the saturation point, when the system is fully reoriented parallel to the field, dispersed CNTs also show homeotropic alignment, contributing their average $\varepsilon_{\parallel}$ ($> \varepsilon_{\perp}$) to the system. Above the saturation ($E_{ac}$ > 80 kV/m) point, the dielectric increment is given by $\Delta\varepsilon_2 = 0.54$. The significant dielectric difference $\Delta\varepsilon = \Delta\varepsilon_2 - \Delta\varepsilon_1 = 0.2$ due to the presence of only 0.005 wt% CNTs in LC media confirms that the dispersed CNTs follow the field-induced director rotation. If the CNTs were to stay in the LC matrix in random orientation without following the nematic director, one would expect $\Delta\varepsilon_1$ to be equal to $\Delta\varepsilon_2$. Thus, parallelly organized CNTs in the nematic matrix can be rotated (between 0° and 90°) mechanically by switching $E_{ac}$ on and off, obtaining directed self-assembly of suspended CNTs. This directed self-assembled system can be

Dielectric relaxation curves for 5CB and 5CB+MWCNT were fitted according to a single-exponential decay function $f(t) = \bar{\varepsilon}_1\, e^{(-t/\tau)} + \bar{\varepsilon}_0$ with a typical regression coefficient of R = 0.9996. Here, $\tau$ is the relaxation decay time, $\bar{\varepsilon}_0$ is the average base dielectric constant, and $\bar{\varepsilon}_1$ is the field-induced average dielectric constant. Thus, the field-saturated average dielectric constant, $\bar{\varepsilon}_{max} = \bar{\varepsilon}_0 + \bar{\varepsilon}_1$. Figure 5a shows the linear dependency of $\bar{\varepsilon}$ with logarithmic time scale. The values for the three fitting parameters, $\bar{\varepsilon}_1$, $\tau$, and $\bar{\varepsilon}_0$ as a function of $E_{ac}$ are shown in Fig. 5b, c, and d respectively. It is obtained from the Fig. 5b that the difference in field-induced average dielectric constant $\bar{\varepsilon}_1$ between 5CB and 5CB+MWCNT after the saturation point is $\Delta\varepsilon = 0.19$. As expected, this value of $\Delta\varepsilon$ has been found to be very close to the value of $\Delta\varepsilon_2 - \Delta\varepsilon_1$ from Fig. 4. It is observed that the relaxation time decreases as $E_{ac}$ increases and, for the composite, saturates at a higher field than that of pure 5CB. This is consistent with the behaviors of $\bar{\varepsilon}_{max}$ shown in Fig. 4. Fig. 5b depicts that, the composite system, for $E_{ac}$ larger than the saturation point ($E_{ac}$ > 80 kV/m) relaxes back faster than pure 5CB. Possibly, dispersed CNTs attract free ions present in the LC media [4]. The presence of ions would slow down the elastic-force driven mechanical relaxation of the nematic domains. The presence of CNTs would lower the free ion concentration allowing the composite system to relax considerably faster. The decrease in the relaxation time is attributed to the decrease in rotational viscosity and increase in anchoring energy [21]. If true, this could be an interesting application for image stabilization in LC displays. The average base dielectric constant $\bar{\varepsilon}_0$ , shown in Fig 5d, does not depend on $E_{ac}$. Also, as expected, the increment in average base dielectric constant $\Delta\bar{\varepsilon}_0$ is found to be equal to $\Delta\varepsilon_1$ from Fig. 4, indicating self-consistency.

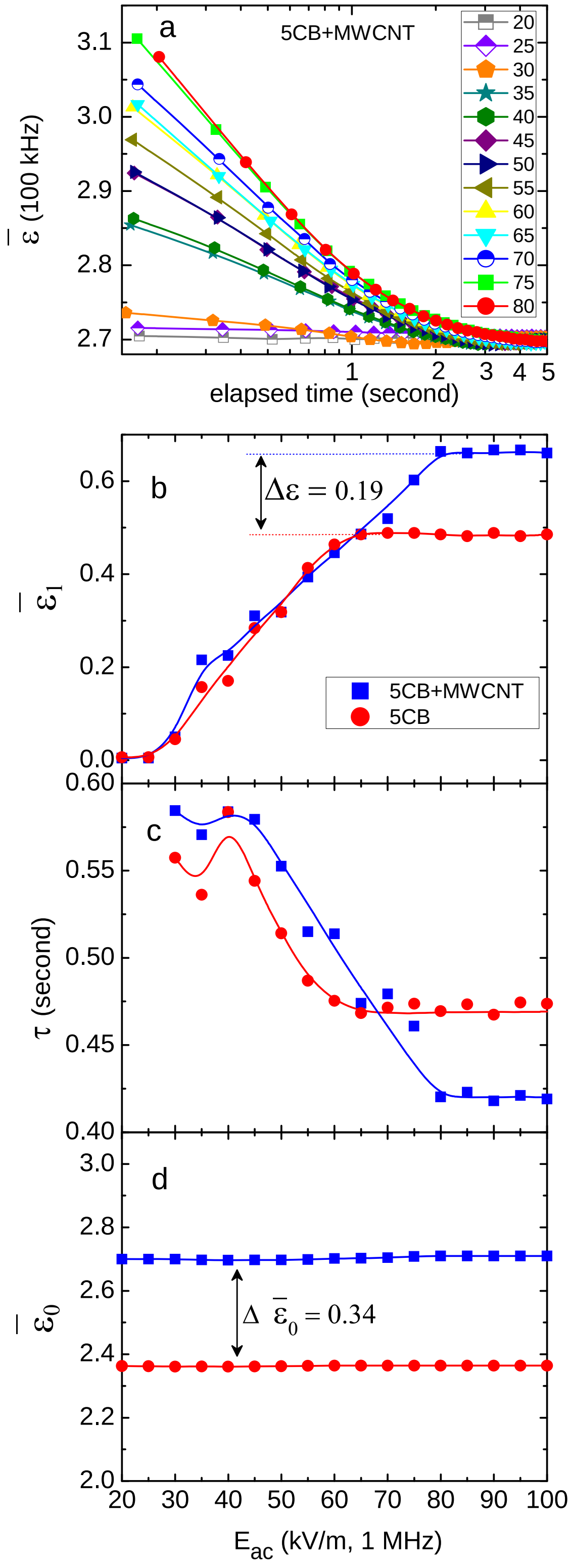

**FIGURE 5: a)** Relaxation dynamic response of $\bar{\varepsilon}$ in logarithmic time scale. The legend represents the magnitude of $E_{ac}$ (1 MHz) in kV/m; **b), c), d)** Fitting parameters according to a single-exponential decay ($f(t) = \bar{\varepsilon}_1 e^{(-t/\tau)} + \bar{\varepsilon}_0$) function for 5CB and 5CB+MWCNT system. Lines represent guide to the eye.

**C. Relaxation dynamics of $\bar{\varepsilon}$ in the isotropic phase**

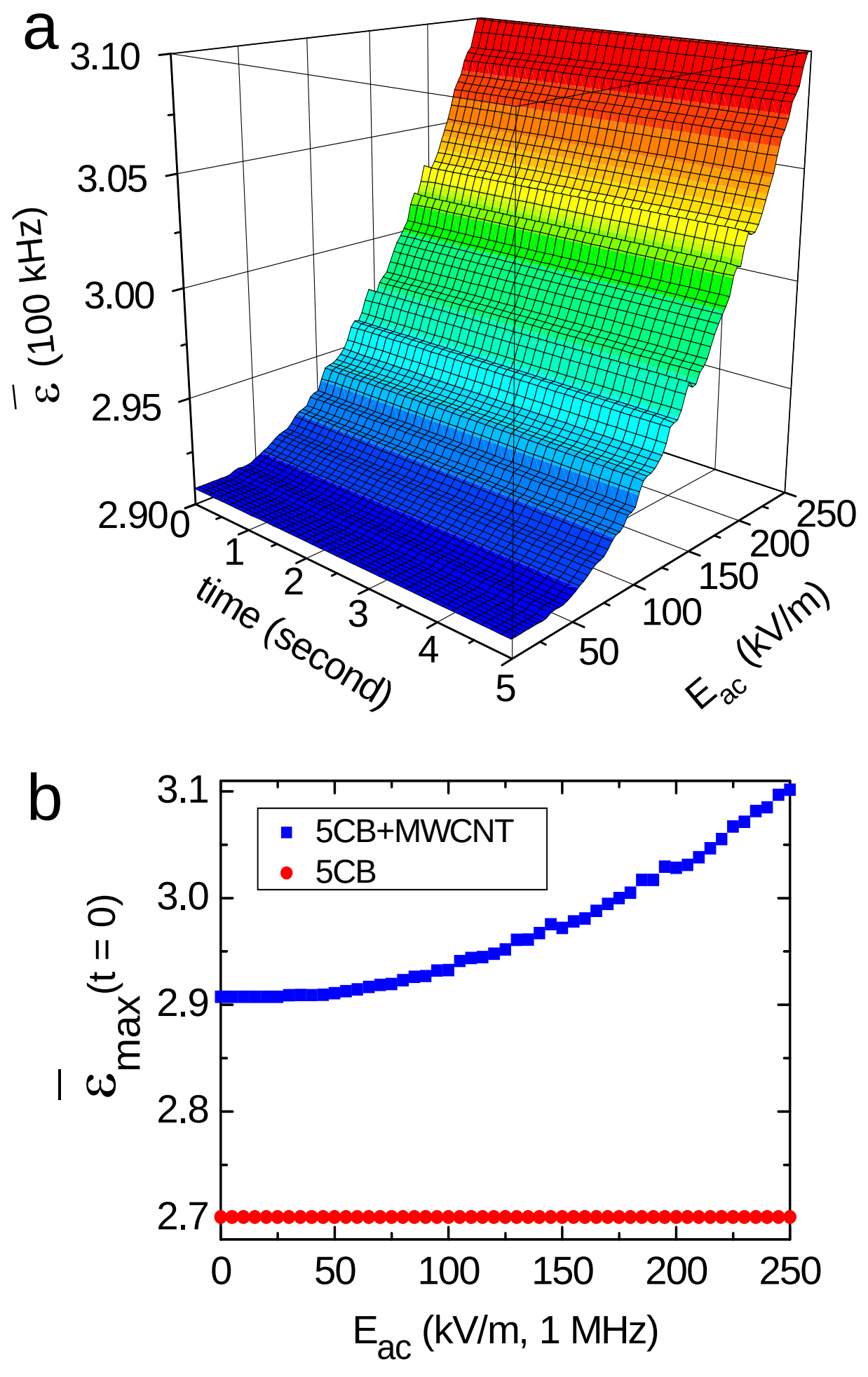

**FIGURE 6: a)** Dynamic response of the average dielectric constant $\bar{\varepsilon}$ for 5CB+MWCNT as a function of *time* and $E_{ac}$ (1 MHz) in the isotropic phase ($T$ = 37°C) after $E_{ac}$ goes off; **b)** Field-saturated dielectric constant, $\bar{\varepsilon}_{max}$ ($\bar{\varepsilon}$ at $t$ = 0) as a function of $E_{ac}$ for 5CB and 5CB+MWCNT in the isotropic phase ($T$ = 37°C). Lines represent guide to the eye.

The same experiment was repeated in the isotropic phase ($T$ = 37 °C) to study the relaxation dynamics. A dramatic change in the field induced orientation mechanism has been observed in this phase. Due to the absence of elastic interactions in the isotropic phase, the LC molecules no longer maintain long range orientation order and act as an isotropic liquid. The isotropic phase, as expected, does not respond to an external field, as also experimentally confirmed in Fig. 6b. But, the composite shows an increment in $\bar{\varepsilon}$ on application of electric field. Interestingly, $\bar{\varepsilon}$ does not relax back over time on switching the field off, as observed in Fig. 6a. This suggests that there is a permanent change in the molecular arrangement each time on application of $E_{ac}$.

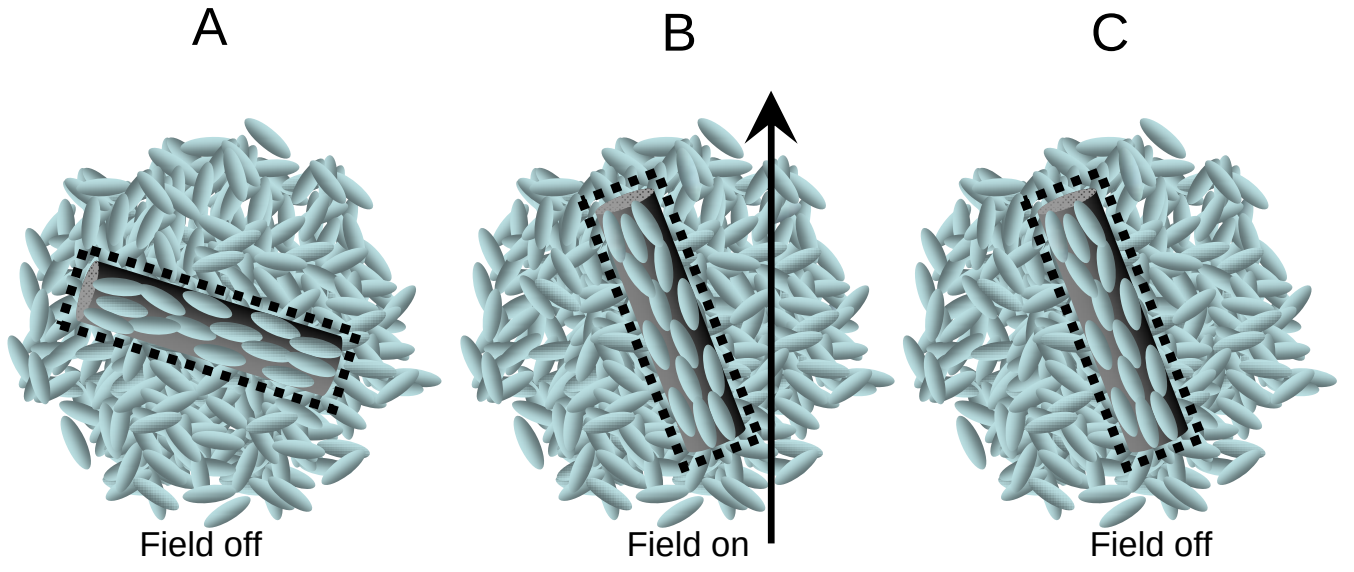

**FIGURE 7:** Schematic diagram of presence of field-responsive anisotropic *pseudo-nematic* domains in the isotropic media. Dashed rectangles represent the LC-CNT *pseudo-nematic* domains

Even though there are no long-range nematic interactions in the isotropic phase, the interaction (surface anchoring) between the LC and CNT surfaces [4,7] still exists. Due to this coupling, the CNT induces local short-range orientation order of LC molecules surrounding the CNT. Presence of few layers of LC molecules on CNT-walls can be visualized as presence of isolated *pseudo-nematic* domains in an isotropic media as described in Fig. 7. These local anisotropic *pseudo-nematic* domains have polarization and they interact with external electric fields. Thus, these field responsive domains can be rotated on application of $E_{ac}$ without disturbing the isotropic media. But, after the field goes off, there is no restoring force in the isotropic LC media to mechanically torque these domains back into the original state. Therefore, each time higher $E_{ac}$ is applied the system goes through some permanent local structural changes, exhibiting a non-volatile memory effect. Figure 6b displays that there is no sharp threshold field to start the reorientation in the isotropic phase and $\bar{\varepsilon}_{max}$ does not seem to saturate in the field range studied. The field induced reorientations of these anisotropic domains can only be erased by slowly cooling the system down to the nematic phase and then heating it up again to the isotropic phase.

**D. Temperature dependence of $\bar{\varepsilon}$ in the nematic and isotropic phase**

For temperature dependent study a different cell was used in order to maintain temperature stabilization of the sample inside the cell. A droplet of each sample was sandwiched between a parallel-plate capacitor configuration, 1 cm diameter and 100 μm spacing, housed in a temperature controlled bath. Dielectric measurements were performed at very low probing field (5kV/m) and at 100 kHz frequency. The normalized $\bar{\varepsilon}$ for 5CB and 5CB+MWCNT are shown in Fig. 8 as a function of temperature shift $\Delta T_{IN}$. The temperature shift is defined as $\Delta T_{IN} = T - T_{IN}$, where $T_{IN}$ is the isotropic ($I$) to nematic ($N$) transition temperature for each sample. The transition temperature is defined as the temperature where $\bar{\varepsilon}$ shows the first discontinuity while entering the $N+I$ phase coexistence region from isotropic phase and was determined from $\bar{\varepsilon}$ vs. $T$ graphs. For 5CB $T_{IN} = 35.1^{o}$C and for 5CB+MWCNT $T_{IN} = 34.67^{o}$C. To compare the dielectric behaviors properly for 5CB and 5CB+MWCNT, the dielectric constants are normalized to the highest temperature (42°C) point studied. The bulk 5CB exhibits the classic temperature dependence of the dielectric constant. Above the transition temperature ($\Delta T_{IN} > 0$) the dielectric constant $\bar{\varepsilon}$ for 5CB flattens out in the isotropic phase and shows no temperature dependence at all. This indicates that the bulk 5CB, for $\Delta T_{IN} > 0$, reaches complete disorder state having order parameter, $S(T) = 0$. The inset in Fig. 8 presents the extracted $d\bar{\varepsilon}/dT$ in the isotropic temperature range. This shows that, $d\bar{\varepsilon}/dT = 0$ for 5CB in the isotropic phase, further indicating the complete isotropic phase of 5CB above the transition temperature.

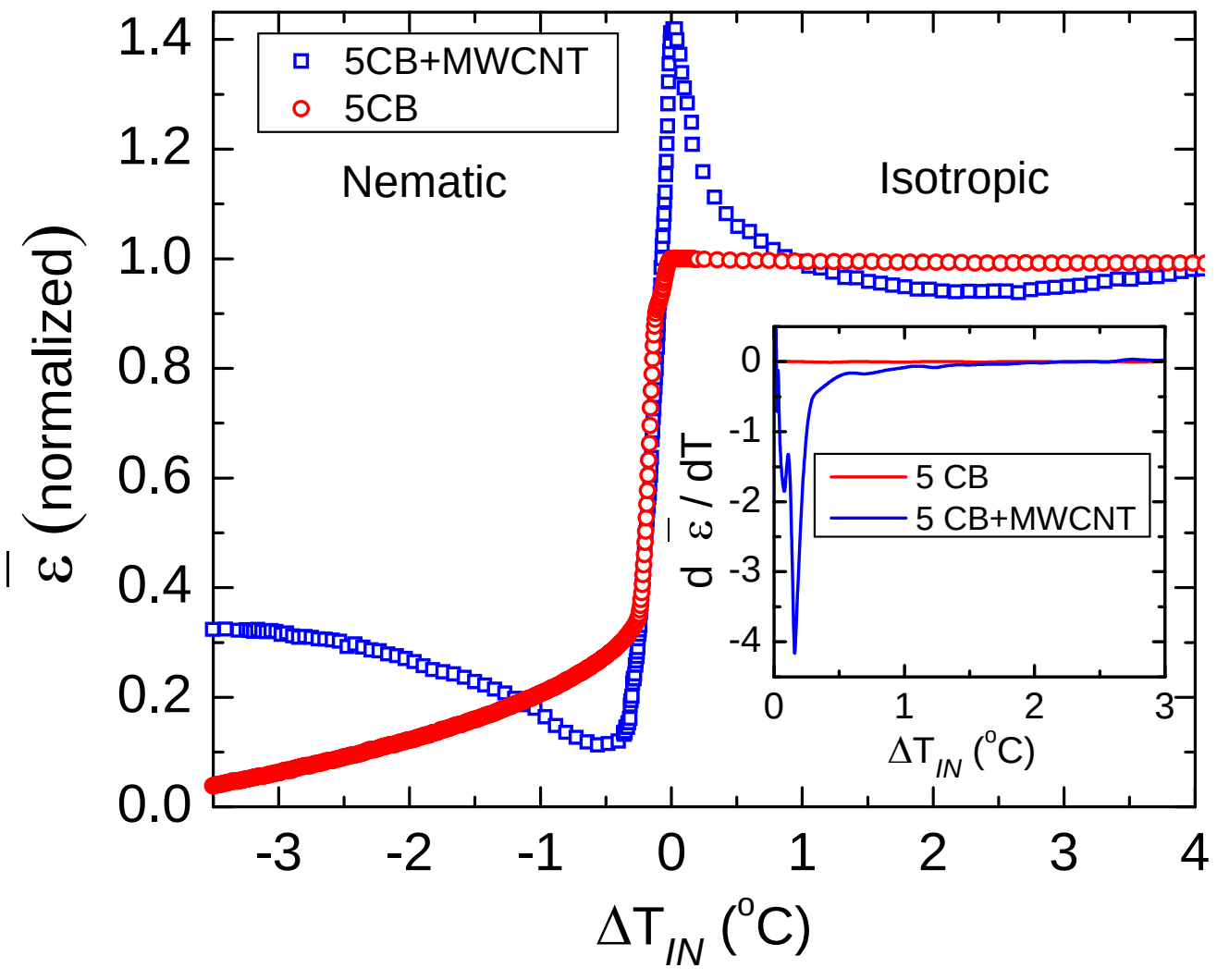

**FIGURE 8:** Normalized average dielectric constant $\bar{\varepsilon}$ for 5CB and 5CB+MWCNT as a function of temperature shift, $\Delta T_{IN} = T - T_{IN}$.

The value of $\bar{\varepsilon}$ for the composite mixture shows a dramatic change both in nematic and isotropic phases, shown in Fig. 8. The presence of CNTs increases the dielectric anisotropy locally in the system due to the anchoring energy. The larger the dielectric anisotropy the smaller the field is needed to make the system respond to it. The evolved wing of $\bar{\varepsilon}$ in the nematic phase ($\Delta T_{IN} < 0$) indicates that the LC+CNT system is more responsive to the low probing field than bulk 5CB; which is another evidence for ferroelectric-type behavior of 5CB+CNT. For this composite system, the curvature in $\bar{\varepsilon}$ and the nonzero value for $d\bar{\varepsilon}/dT$ in the isotropic phase ($\Delta T_{IN} > 0$) imply that the system does not reach a complete disorder state (order parameter, $S(T) \neq 0$), indicating the presence of *pseudo-nematic* domains discussed earlier.

## IV. CONCLUSIONS

We have demonstrated the dielectric hysteresis, relaxation response, and nematic to isotropic phase transition phenomena for a LC-CNT hybrid composite by probing its dielectric properties. The field annealing shows ferroelectric-type hysteresis effect in the composite system, indicating an improvement in the nematic order. The presence of a low concentration well-dispersed CNTs causes an increase in the spontaneous polarization in the nematic matrix due to the strong anchoring energy associated with the alignment mechanism. The dielectric relaxation dynamics reveal that incorporating CNTs in a nematic platform results in an improvement in relaxation decay time (for $E_{ac}$ > 80 kV/m), decreasing rotation viscosity. The local anisotropic *pseudo-nematic* domains in the isotropic phase demonstrate a field-induced non-volatile memory effect. The dielectric constant as a function temperature represents *I–N* phase transition phenomena for the bulk and the composite. A strong temperature dependent dielectric constant for the LC+CNT system in the isotropic phase confirms the presence of *pseudo-nematic* domains. This versatile nano-scale electro-mechanical system might reveal interesting hysteresis effect in the high frequency regime due to their high frequency switching effect. Future work involves frequency dependent dielectric-hysteresis studies for different CNT concentrations in LC media for both the nematic and isotropic phases.

**ACKNOWLEDGEMENTS:**
The authors are grateful to Rafael Garcia for useful discussions. This work was supported by the Department of Physics, WPI.